\documentclass[12pt]{article}
\usepackage{pifont}
\usepackage{graphicx}
\usepackage{amsmath}
\usepackage{amssymb}
\usepackage{amsfonts}
\usepackage[numbers,sort&compress]{natbib}
\title{Cusp solitons to the long-short waves equation and the $\bar\partial$-dressing method
\footnotetext{}}
\author{Junyi Zhu\thanks{Email: jyzhu@zzu.edu.cn} and Yonghui Kuang\\
{\small School of Mathematics and Statistics, Zhengzhou University,}\\
{\small Zhengzhou, Henan 450001,
People's Republic of China}}
\date{}
\begin{document}
\maketitle
\begin{abstract}
The dressing method based on $3\times3$ matrix $\bar{\partial}$-problem is extended to study the long-short waves equation with the cases $\sigma=\pm1$.
The soliton solutions for the long-short wave equation for $\sigma=\pm1$ are given by means of the properties of Cauchy matrix.
It is shown that the long-short wave equation for $\sigma=1$ has the cusp solitons.
\end{abstract}
\section{Introduction}
The purpose of this paper is to study the long-short waves (LSW) equation, which reads
\begin{equation}\label{a1}
\begin{aligned}
&u_t=\mathrm{i}u_{xx}-v_{x}u+\mathrm{i}{v^2}u-2\mathrm{i}{\sigma}u|u|^2,\\
&v_t=2{\sigma}(|u|^2)_{x},\quad \sigma=\pm1,
\end{aligned}
\end{equation}
where $u=u(x,t)$ represents the envelope of the short wave and $v=v(x,t)$ represents the amplitude of the long wave.
As an integrable system, this equation has been widely studied by many methods such as inverse scattering transformation,
Painlev\'{e} analysis, B\"{a}cklund and Darboux transformation \cite{Newell,Newell A,Chowdhury A R,Chowdhury A,Ling}.
We note that long wave-short wave resonance can be achieved in a second-order nonlinear negative refractive index medium when the short wave
lies on the negative index branch \cite{Chowdhury}.
In \cite{Liu}, it was shown that (\ref{a1}) is associated with a model equation proposed by Yajima and Oikawa through a Muira transformation.

The $\bar{\partial}$-dressing method \cite{A-B-F,Zakharov,Bogdanov,Beals,Zakharov V,Santini} is a powerful tools to construct and solve integrable nonlinear equations,
moreover, it can also describes their transformations and reductions \cite{Konopelchenko,Doktorov}.
In this paper, we will consider the LSW equation in virtue of extended $\bar{\partial}$-dressing method \cite{Zhu,Z-G2} and give their $N$-soliton solution.
In Sect.2, Using the $\bar{\partial}$-dressing method,
we obtain the Lax pair for the LSW equation . In Sect.3, suitable symmetry
conditions are applied to derive the LSW equation in both cases $\sigma=\pm1$. In Sect.4, The soliton solutions for the LSW equation for $\sigma=\pm1$ are given,
moreover, the explicit expression of $N$-soliton solution is found for this equation by means of the properties of Cauchy matrix.
We note that for the LSW equation with $\sigma=1$, we obtain the one and two cusp solitons which are singular spiky soliton solutions \cite{W-I-S}.

\setcounter{equation}{0}
\section{Spectral transform and Lax pair}
In this section, we consider the $3\times3$ matrix $\bar{\partial}$-problem in the complex $k$-plane
\begin{equation}\label{b1}
\bar{\partial}\psi(k,\bar{k})=\psi(k,\bar{k})R(k,\bar{k}),
\end{equation}
where $R=R(k,\bar{k})$ is a spectral transform matrix associated with a nonlinear equation.
It is easy to see that a solution of the $\bar{\partial}$-problem (\ref{b1}) with the canonical normalization
can be written as
\begin{equation}\label{b2}
\psi(k)=I+\psi RC_k,
\end{equation}
where $C_k$ denotes the Cauchy-Green integral operator acting on the left
$$\psi RC_k=\frac{1}{2\mathrm{i}\pi}\iint\frac{{\rm d}z\wedge {\rm d}\bar{z}}{z-k}\psi(z)R(z),$$
and here we have suppressed the variable $\bar{k}$ dependence in $\psi$ and $R$.

The aim of the $\bar{\partial}$-dressing method is to construct the compatible system of linear equations for $\psi$ and the
nonlinear evolution equations associated with the $\bar{\partial}$-problem (\ref{b1}). According to the idea of the inverse scattering transform method,
it is important to introduce the $x,t$ dependence in the spectral transform matrix $R(k,\bar{k})$.
For the LSW equation (\ref{a1}), we introduce the following linear system about the spectral transform matrix as
\begin{equation}\label{b14}
R_x=\mathrm{i}k[J,R],\quad J={\rm diag}(1,0,-1),
\end{equation}
and
\begin{equation}\label{b21}
R_t=\mathrm{i}k^2[\tilde{J},R], \quad\tilde{J}={\rm diag}\left(-\frac{1}{3}, \frac{2}{3},-\frac{1}{3}\right).
\end{equation}

Following the extended $\bar\partial$-dressing method in \cite{Zhu,Z-G2}, we have
the following linear spectral problem
\begin{equation}\label{b19}
\psi_x-\mathrm{i}k[J,\psi]=Q\psi, \quad Q=\mathrm{i}[J,\langle\psi R\rangle],
\end{equation}
and
\begin{equation}\label{b27}
\psi_t=\mathrm{i}k^2[\tilde{J},\psi]-k(JQ+QJ)\psi+[\mathrm{i}Q^2-(J\langle\psi R\rangle_x+\langle\psi R\rangle_xJ)]\psi,
\end{equation}
where
$$\langle\psi R\rangle=\frac{1}{2i\pi}\iint \psi(k)R(k){\rm d}k\wedge{\rm d}\bar{k}.$$

Our next task is to establish a relation between the representation $J\langle\psi R\rangle_x+\langle\psi R\rangle_xJ$ in (\ref{b27})
and the potential $Q$. To this end, we first need to introduce a symmetry condition about $Q$ for obtaining the LSW equation
\begin{equation}\label{b28}
\mathcal{A}Q\mathcal{A}=Q,\quad \mathcal{A}=\left(\begin{matrix}
0&0&1\\
0&1&0\\
1&0&0
\end{matrix}\right),
\end{equation}
which implies that $Q$ takes the form
\begin{equation}\label{b29}
Q=\left(\begin{matrix}
0&u&\mathrm{i}v\\
w&0&w\\
\mathrm{i}v&u&0
\end{matrix}\right)=Q^{(s)}+Q^{(o)},
\end{equation}
where we have use the notation for a $3\times3$ matrix $A=A^{(d)}+A^{(s)}+A^{(o)}$ with
$$A^{(d)}=\left(\begin{matrix}
A_{11}&0&0\\
0&A_{22}&0\\
0&0&A_{33}
\end{matrix}\right),\quad A^{(s)}=\left(\begin{matrix}
0&A_{12}&0\\
A_{21}&0&A_{23}\\
0&A_{32}&0
\end{matrix}\right),\quad A^{(o)}=\left(\begin{matrix}
0&0&A_{13}\\
0&0&0\\
A_{31}&0&0
\end{matrix}\right).$$

It is readily verified that if $\psi$ is the solution of the spectral problem (\ref{b19}) then $\psi R$ satisfies
the same problem. From the latter equation, we have
\begin{equation}\label{b25}
\langle\psi R\rangle_x=\mathrm{i}[J,\langle k\psi R\rangle]+Q\langle\psi R\rangle.
\end{equation}
Then from (\ref{b19}) and the diagonal part of equation (\ref{b25}), we find
\begin{equation}\label{b30}
\begin{aligned}
\langle\psi R\rangle^{(s)}&=-\mathrm{i}[J,Q^{(s)}],\quad \langle\psi R\rangle^{(o)}=-\frac{\mathrm{i}}{4}[J,Q^{(o)}], \\
&\langle\psi R\rangle_x^{(d)}=\mathrm{i}(uw-\frac{1}{2}v^2)J.
\end{aligned}
\end{equation}
Thus, the time-dependent linear equation is given by
\begin{equation}\label{b31}
\psi_t=\mathrm{i}k^2[\tilde{J},\psi]-k(JQ+QJ)\psi+[\mathrm{i}Q^2+\mathrm{i}[J^2,Q_x^{(s)}]-\mathrm{i}(2uw-v^2)J^2]\psi.
\end{equation}

\setcounter{equation}{0}
\section{Long-short wave equation}
In order to derive the LSW equation, we differentiate $Q$ in (\ref{b19}) with respect to $t$,
and use the properties of the $\bar\partial$-problem (\ref{b1}), then \cite{Doktorov,Zhu,Z-G2}
\begin{equation}\label{c1}
Q_t=\mathrm{i}[J,\langle\psi R\rangle_t]=[J,\langle\bar\partial(k^2U)\rangle],
\end{equation}
where $U=U(x,t,k)$ is defined as
\begin{equation}\label{c4}
U=\psi\tilde{J}\psi^{-1}.
\end{equation}
Furthermore, From the linear spectral problem (\ref{b19}), we find
\begin{equation}\label{c5}
U_x=\mathrm{i}k[J,U]+[Q,U].
\end{equation}

On the other hand, equations (\ref{c4}) and (\ref{b2}) imply that $U$ has the following asymptotic expansion
\begin{equation}\label{c6}
U=\tilde{J}+\sum\limits_{j=1}^\infty\frac{1}{k^j}u_j(x,t),\quad k\rightarrow\infty.
\end{equation}
Then equation (\ref{c1}) reduces to
\begin{equation}\label{c7}
\begin{aligned}
Q_t=&[J,\langle\bar\partial(k^2\tilde{J}+\sum\limits_{j=1}^\infty u_j(x,t)k^{2-j})\rangle]\\
=&[J,\langle\sum\limits_{j=1}^\infty u_j(x,t)\pi\delta(k)\delta_{j,3}\rangle]\\
=&-[J,u_3(x,t)],
\end{aligned}
\end{equation}
in view of the identity $\bar\partial k^{n-j}=\pi\delta(k)\delta_{j,n+1}$.
Now, substitution of the expansion (\ref{c6}) into (\ref{c5}), we observe that
\begin{subequations}
\begin{equation}\label{c8a}
\mathrm{i}[J,u_1]=[\tilde{J},Q],
\end{equation}
\begin{equation}\label{c8b}
u_{1,x}=\mathrm{i}[J,u_2]+[Q,u_1],
\end{equation}
\begin{equation}\label{c8c}
u_{2,x}=\mathrm{i}[J,u_3]+[Q,u_2],
\end{equation}
$$\cdots$$
\end{subequations}
From equation (\ref{c8a}), we know that
\begin{equation}\label{c9}
u_1^{(o)}=0,\quad u_1^{(s)}=-\mathrm{i}[J,[\tilde{J},Q].
\end{equation}
Moreover, equation (\ref{c8b}) implies that $u_1^{(d)}=0$ and
\begin{equation}\label{c10}
\begin{aligned}
u_{1,x}^{(s)}-[Q^{(o)},u_1^{(s)}]=\mathrm{i}[J,u_2^{(s)}],\\
[Q^{(s)},u_1^{(s)}]=-\mathrm{i}[J,u_2^{(o)}].
\end{aligned}
\end{equation}
Hence,
\begin{equation}\label{c11}
\begin{aligned}
&u_2^{(s)}=-\mathrm{i}[J,u_{1,x}^{(s)}]+\mathrm{i}[J,[Q^{(o)},u_1^{(s)}]],\\
&u_2^{(o)}=\frac{\mathrm{i}}{4}[J,[Q^{(s)},u_1^{(s)}]].
\end{aligned}
\end{equation}
In addition, from (\ref{c7}) and (\ref{c8c}), we obtain
\begin{equation}\label{c12}
Q_t=\mathrm{i}\left(u_{2,x}-[Q,u_2]\right),
\end{equation}
which implies that $u_2^{(d)}=3uw\tilde{J}$ and then the coupled equations can be obtained
\begin{equation}\label{c13}
\begin{aligned}
u_t&=\mathrm{i}u_{xx}-uv_x-2\mathrm{i}u^2w+\mathrm{i}uv^2,\\
w_t&=-\mathrm{i}w_{xx}-wv_x+2\mathrm{i}uw^2-\mathrm{i}wv^2,\\
v_t&=2(uw)_x.
\end{aligned}
\end{equation}
We note that equations (\ref{c13}) reduces to the LSW equation, if $w=\sigma\bar{u}$ or $Q^\dagger=-\mathcal{B}Q\mathcal{B},~\mathcal{B}={\rm diag}\{1,-\sigma,1\}$, we can obtain (\ref{a1}).
For the purpose of finding the solution of the obtained equations, we need other symmetry conditions
\begin{equation}\label{c14}
\mathcal{A}\psi(k)\mathcal{A}=\psi(-k),\quad \psi^\dagger(\bar{k})=\mathcal{B}\psi^{-1}\mathcal{B},
\end{equation}
where the matrix $\mathcal{A}$ is defined in (\ref{b28}).
\setcounter{equation}{0}
\section{Solutions and solitons}
In this section, we will give the explicit solutions of the LSW equations. To this end, we
introduce the following spectral transform matrix as
\begin{equation}\label{d1}
R=\pi\sum\limits_{j=1}^N{\begin{pmatrix}
0&c_je^{-\mathrm{i}k^2t}\delta(k-k_j)&0\\
d_je^{\mathrm{i}k^2t}\delta(k-l_j)&0&-d_je^{\mathrm{i}k^2t}\delta(k+l_j)\\
0&-c_je^{-\mathrm{i}k^2t}\delta(k+k_j)&0
\end{pmatrix}},
\end{equation}
where the constants $k_j,l_j$ are mutually distinct and the functions $c_j=c_j(x)$ and $d_j=d_j(x)$ satisfy the equation
\begin{equation}\label{d2}
c_{j,x}=\mathrm{i}k_jc_j,\quad d_{j,x}=-\mathrm{i}l_jd_j.
\end{equation}
Then, from (\ref{b19}), the solution of equation (\ref{c13}) is given by
\begin{equation}\label{d3}
\begin{aligned}
&u=\mathrm{i}\langle\psi R\rangle_{12}=-i\hat\phi g^T,\\
&w=\mathrm{i}\langle\psi R\rangle_{23}=\mathrm{i}\tilde\psi_{22}h^T,\\
&v=2\langle\psi R\rangle_{13}=2\tilde\psi_{12}h^T,
\end{aligned}
\end{equation}
where
\begin{equation}\label{d4}
\begin{aligned}
&\hat\phi=(\phi(k_1),\phi(k_2),\cdots,\phi(k_N)),\qquad \phi(k_j)=\psi_{11}(k_j)-\psi_{13}(-k_j),\\
&\tilde\psi_{22}=(\psi_{22}(-l_1),\cdots,\psi_{22}(-l_N)),\quad \tilde\psi_{12}=(\psi_{12}(-l_1),\cdots,\psi_{12}(-l_N)),\\
\end{aligned}
\end{equation}
and
\begin{equation}\label{d5}
\begin{aligned}
&g=(g_1,\cdots,g_N),\quad g_j=c_je^{-\mathrm{i}k^2t}=e^{\mathrm{i}(k_jx-k_j^2t+\xi_j)},\\
&h=(h_1,\cdots,h_N),\quad h_j=d_je^{\mathrm{i}l_j^2t}=e^{-\mathrm{i}(l_jx-l_j^2t+\eta_j)},
\end{aligned}
\end{equation}
with $\xi_j,\eta_j$ are arbitrary constants. Then we need to give the representations of the vectors in (\ref{d4}) about the discrete data.
To this end, substituting (\ref{d1}) into (\ref{b2}) yields
\begin{equation}\label{d6}
\begin{aligned}
&\psi_{11}(k)=1-\sum\limits_{j=1}^N\frac{h_j\psi_{12}(l_j)}{l_j-k},\\
&\psi_{12}(k)=-\sum\limits_{j=1}^N\left(\frac{g_j\psi_{11}(k_j)}{k_j-k}+\frac{g_j\psi_{13}(-k_j)}{k_j+k}\right),\\
&\psi_{13}(k)=-\sum\limits_{j=1}^N\frac{h_j\psi_{12}(-l_j)}{l_j+k},
\end{aligned}
\end{equation}
which implies that
\begin{equation}\label{d7}
\begin{aligned}
\phi(k_j)&=1+\sum\limits_{n=1}^N\sum\limits_{m=1}^N\phi(k_n)\left(\frac{g_n}{k_n+l_m}-\frac{g_n}{k_n-l_m}\right)\frac{h_m}{l_m-k_j},\\
\psi_{12}(-l_j)&=-\sum\limits_{n=1}^N\frac{g_n}{k_n+l_j}+\sum\limits_{m=1}^N\sum\limits_{n=1}^N\psi_{12}(l_m)\frac{h_m}{l_m-k_n}\frac{g_n}{k_n+l_j}\\
&\quad+\sum\limits_{m=1}^N\sum\limits_{n=1}^N\psi_{12}(-l_m)\frac{h_m}{l_m-k_n}\frac{g_n}{k_n-l_j},\\
\psi_{12}(l_j)&=-\sum\limits_{n=1}^N\frac{g_n}{k_n-l_j}+\sum\limits_{m=1}^N\sum\limits_{n=1}^N\psi_{12}(l_m)\frac{h_m}{l_m-k_n}\frac{g_n}{k_n-l_j}\\
&\quad+\sum\limits_{m=1}^N\sum\limits_{n=1}^N\psi_{12}(-l_m)\frac{h_m}{l_m-k_n}\frac{g_n}{k_n+l_j},\\
\end{aligned}
\end{equation}
Let us introduce a set of matrix
$$\begin{aligned}
&G=(G_{nj})=\left(\frac{g_n}{k_n-l_j}\right)_{N\times N},\quad \tilde{G}=(\tilde{G}_{nj})=\left(\frac{g_n}{k_n+l_j}\right)_{N\times N},\\
&H=(H_{mn})=\left(\frac{h_m}{l_m-k_n}\right)_{N\times N},\quad \hat\psi_{12}=(\psi_{12}(l_1),\cdots,\psi_{12}(l_N)),\\
&{\mathcal{G}}=({\mathcal{G}}_1,\cdots,{\mathcal{G}}_N),\quad {\mathcal{G}}_j=\sum\limits_{n=1}^N\frac{g_n}{k_n-l_j},\\
&\tilde{\mathcal{G}}=(\tilde{\mathcal{G}}_1,\cdots,\tilde{\mathcal{G}}_N),\quad \tilde{\mathcal{G}}_j=\sum\limits_{n=1}^N\frac{g_n}{k_n+l_j}
\end{aligned}$$
Then equation (\ref{d7}) is written as
\begin{equation}\label{d8}
\begin{aligned}
&\hat\phi=E-\hat\phi(\tilde{G}-G)H,\\
&\tilde{\psi}_{12}=-\tilde{\mathcal{G}}+\hat\psi_{12}H\tilde{G}+\tilde\psi_{12}HG,\\
&\hat\psi_{12}=-{\mathcal{G}}+\hat\psi_{12}HG+\tilde\psi_{12}H\tilde{G},
\end{aligned}
\end{equation}
where $E=(1,\cdots,1)$ denotes a $n$-dimensional vector. From (\ref{d8}) and (\ref{d3}), we know that the explicit solution of LSW equation takes the form
\begin{equation}\label{d9}
\begin{aligned}
&u=-\mathrm{i}E(I+\tilde{M})^{-1}g^T,\\
&v=[(\mathcal{G}-\tilde{\mathcal{G}})(I+M)^{-1}-(\mathcal{G}+\tilde{\mathcal{G}})(I-N)^{-1}]h^T,
\end{aligned}
\end{equation}
where
$$\begin{aligned}
&\tilde{M}=(\tilde{G}-G)H,\\
&M=H(\tilde{G}-G),\\
&N=H(\tilde{G}+G).
\end{aligned}$$
It is noted that the explicit representation of $w$ can be obtained in a same way. One finds
$$\begin{aligned}
&\tilde\psi_{22}=E+\hat\psi_{22}H\tilde{G}+\tilde\psi_{22}HG,\\
&\hat\psi_{22}=E+\hat\psi_{22}HG+\tilde\psi_{22}H\tilde{G},
\end{aligned}$$
which imply that $\tilde\psi_{22}=\hat\psi_{22}$, then
\begin{equation}\label{d10}
w=\mathrm{i}E(I-N)^{-1}h^T.
\end{equation}
We note that for the LSW equation, $w=\sigma{\bar{u}}$. Hence $l_j=\bar{k_j},\ d_j=\sigma{\bar{c_j}}$  and
\begin{equation}\label{d11}
\begin{aligned}
&g_j=e^{-z_j+\mathrm{i}\varphi_j},\quad h_j=\sigma\bar{g}_j, \quad k_j=\xi_j+\mathrm{i}\eta_j,\\
&z_j=\eta_jx-2\xi_j\eta_jt+z_0,\quad \varphi_j=\xi_jx-(\xi_j^2-\eta_j^2)t+\varphi_0,
\end{aligned}
\end{equation}
where $z_0$ and $\varphi_0$ are arbitrary real constants.
It is remarked that this condition can also be obtained by means of the symmetry conditions (\ref{c14}).
From (\ref{d9}), we have
\begin{equation}\label{d12}
\begin{aligned}
u=&-\mathrm{i}{\rm tr}[(I+\tilde{M})^{-1}{g^T}E]\\
=&-\mathrm{i}\frac{\det(I+\tilde{M}+{g^T}E)-\det(I+\tilde{M})}{\det(I+\tilde{M})}\\
v=&{\rm tr}[(I+M)^{-1}{h^T}(\mathcal{G}-\tilde{\mathcal{G}})]-{\rm tr}[(I-N)^{-1}{h^T}(\mathcal{G}+\tilde{\mathcal{G}})]\\
=&\frac{\det(I+M+{h^T}(\mathcal{G}-\tilde{\mathcal{G}}))-\det(I+M)}{\det(I+M)}\\
&-\frac{\det(I-N+{h^T}(\mathcal{G}+\tilde{\mathcal{G}}))-\det(I-N)}{\det(I-N)}.
\end{aligned}
\end{equation}

In the following, we will give the one-soliton and two-soliton solutions. Above all, for the case of $N=1$
\begin{equation}\label{d13}
\begin{aligned}
&u=-\mathrm{i}g_1\left(1+\frac{2\sigma\bar{k}_1|g_1|^2}{({\bar{k}_1}-{k_1})^2(\bar{k}_1+k_1)}\right)^{-1}\\
&v=-\frac{2\sigma|g_1|^2}{\bar{k_1}+k_1}\left|1+\frac{2\sigma{\bar{k_1}}|g_1|^2}{({\bar{k}_1}-{k_1})^2(\bar{k_1}+k_1)}\right|^{-2}.
\end{aligned}
\end{equation}
Then, one soliton solution of LSW equation takes the form
\begin{equation}\label{d14}
\begin{aligned}
&u=\frac{-\mathrm{i}}{D}(e^{2\vartheta-z_1+\mathrm{i}\varphi_1}-\sigma e^{-z_1+\mathrm{i}(\varphi_1+{\rm arg}k_1)}),\quad v=-\sigma\frac{\alpha}{D},\\
&D=e^{2\vartheta}+e^{-2\vartheta}-\sigma\frac{\bar{k}_1+k_1}{|k_1|},\quad \alpha=\frac{-(\bar{k}_1-k_1)^2}{|k_1|},
\end{aligned}
\end{equation}
where $\vartheta$ is defined by
$$-\frac{2|g_1|^2|k_1|}{(\bar{k}_1-k_1)^2(\bar{k}_1+k_1)}=e^{-2\vartheta}.$$
Figure 1 and 2 describe the one-soliton solution for $\sigma=-1$ and $\sigma=1$, respectively.
\begin{figure}[h]
\centering
\includegraphics[width=4.0cm,angle=-90]{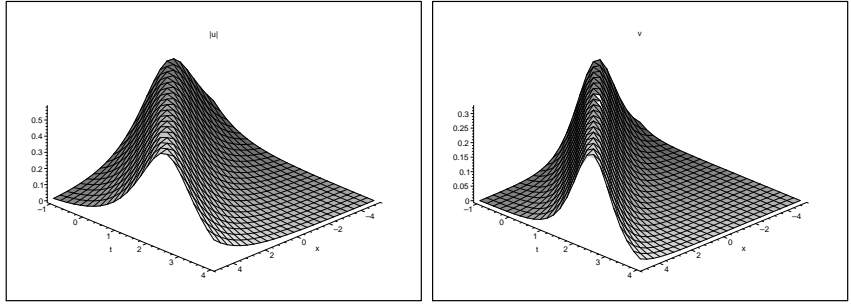}
\includegraphics[width=4.0cm,angle=-90]{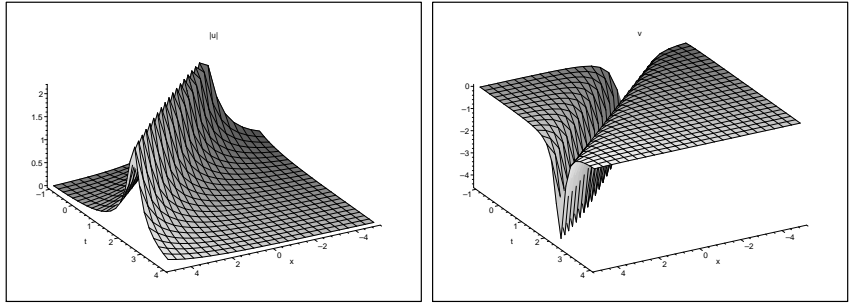}
\caption{$k_1=1.04+0.6\mathrm{i},\sigma=-1,z_0=0,\varphi_0=0$.}
\end{figure}
\begin{figure}[h]
\centering
\includegraphics[width=4.0cm,angle=-90]{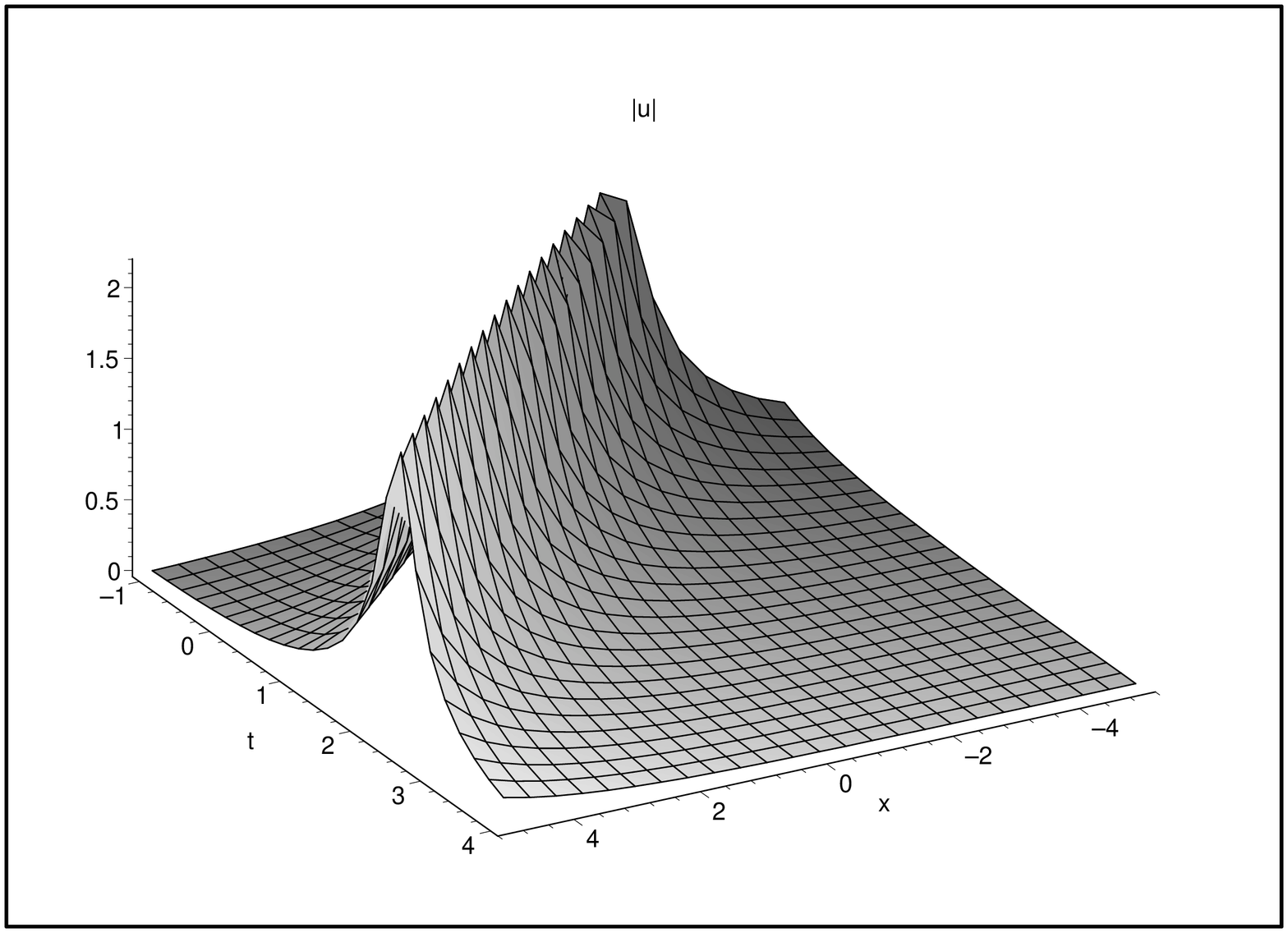}
\includegraphics[width=4.0cm,angle=-90]{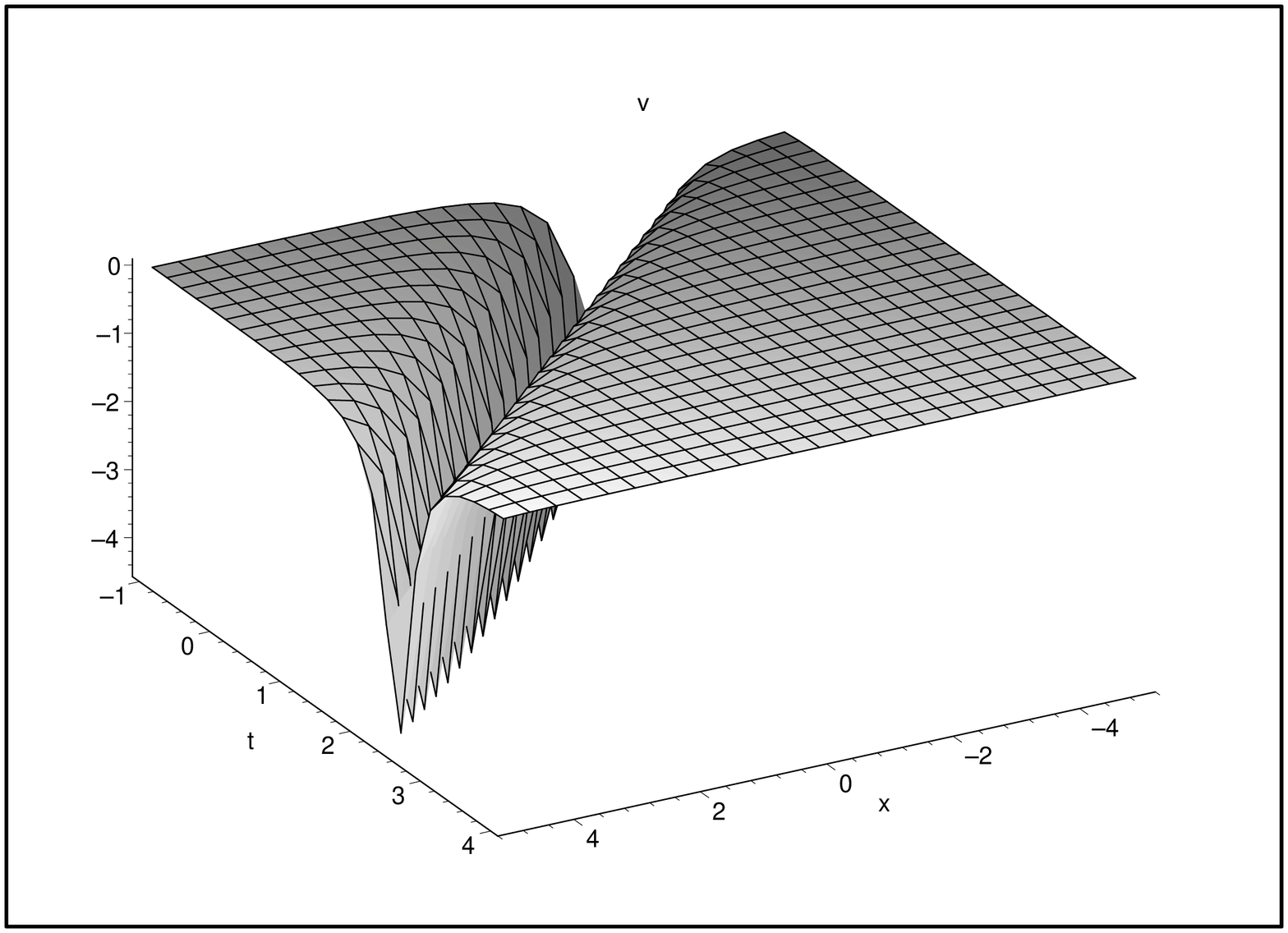}
\caption{$k_1=1.04+0.6\mathrm{i},\sigma=1,z_0=0,\varphi_0=0$.}
\end{figure}

Now, we will derive the explicit $N$-soliton solution of (\ref{a1}). By virtue of the method of linear algebra, we know that
\begin{equation}\label{d18}
\det(I+\tilde{M})=1+\sum_{\nu=1}^N\sum_{1\leq{j_1}<{j_2}<\ldots<{j_{\nu}}\leq{N}}\tilde{M}\begin{pmatrix}j_1&j_2&\ldots&j_{\nu}\end{pmatrix},
\end{equation}
where $\tilde{M}\begin{pmatrix}j_1&j_2&\ldots&j_{\nu}\end{pmatrix}$ denotes the principal minor of $N\times{N}$ matrix $\tilde{M}$ obtained by taking all the elements of $(j_1,j_2,\ldots,j_{\nu})-th$ columns and rows. By using the Cauchy-Binet formula, we can calculate the value of $\tilde{M}\begin{pmatrix}j_1&j_2
&\ldots&j_{\nu}\end{pmatrix}$
\begin{equation}\label{d19}
\tilde{M}\begin{pmatrix}j_1&j_2&\ldots&j_{\nu}\end{pmatrix}=\sum_{1\leq{r_1}<\ldots<{r_{\nu}}\leq{N}}\tilde{H}\begin{pmatrix}j_1&j_2&\ldots&j_{\nu}\\r_1&
r_2&\ldots&r_{\nu}\end{pmatrix}H\begin{pmatrix}r_1&r_2&\ldots&r_{\nu}\\j_1&j_2&\ldots&j_{\nu}\end{pmatrix}
\end{equation}
where $\tilde{H}=(\tilde{G}-G)H$ and $\tilde{H}\begin{pmatrix}j_1&j_2&\ldots&j_{\nu}\end{pmatrix}$ denotes
the determinant of the submatrix obtained by preserving the $(j_1,j_2,\ldots,j_{\nu})-th$ rows
and $(r_1,r_2,\ldots,r_{\nu})-th$ columns of $\tilde{H}$; $H\begin{pmatrix}\cdot\\\cdot\end{pmatrix}$ denotes similarly the determinant of the submatrix for $H$.
It is noted that $\tilde{H}$ is Cauchy type matrices,  then
we obtain the explicit representation of $\det(I+\tilde{M})$ 
\begin{equation}\label{d21}
\begin{aligned}
\tilde{\mathcal{D}}&=\det(I+\tilde{M})\\
&=1+\sum_{\nu=1}^N\sum_{1\leq{j_1}<{j_2}<\ldots<{j_{\nu}}\leq{N}}\sum_{1\leq{r_1}<{r_2}<\ldots<{r_{\nu}}\leq{N}}(-2)^{\nu}\prod\limits_{l,m}\bar{k}_mg_lh_m\\
&\quad\times\prod_{l<l^{\prime},m<m^{\prime}}\frac{({k_l}^2-{k_{l^{\prime}}}^2)(k_l-k_{l^{\prime}})
({\bar{k}_m}^2-{\bar{k}_{m^{\prime}}}^2)(\bar{k}_m-\bar{k}_{m^{\prime}})}{({k_l}^2-{\bar{k}_m}^2)(k_l-\bar{k}_m)},
\end{aligned}
\end{equation}
where $m,m^{\prime}\in\{r_1,r_2,\ldots,r_{\nu}\}$ and $l,l^{\prime}\in\{j_1,j_2,\ldots,j_{\nu}\}$.
At the same time, it is readily verified that
\begin{equation}\label{d22}
\det(I+\tilde{M})=\det(I+M).
\end{equation}
In the following, we will evaluate the numerator of the expressions in (\ref{d12}). To this end, let
\begin{equation}\label{d23}
C=\tilde{M}+{g^T}E=\tilde{K}K,
\end{equation}
where $\tilde{K}=\begin{pmatrix}g^T,&\tilde{H}\end{pmatrix}=(\tilde{K}_{nm})$ and $K=\begin{pmatrix}E\\
H\end{pmatrix}=(K_{mn})$, with $n\in\{1,2,\ldots,N\}$,
$m\in\{0,1,2,\ldots,N\}$. Hence, $\det(I+C)$ takes the same expansion as (\ref{d18}), where
\begin{equation}\label{d24}
C\begin{pmatrix}j_1&j_2&\ldots&j_{\nu}\end{pmatrix}=\sum_{1\leq{r_1}<\ldots<{r_{\nu}}\leq{N}}\tilde{K}\begin{pmatrix}j_1&j_2&\ldots&j_{\nu}\\r_1&
r_2&\ldots&r_{\nu}\end{pmatrix}K\begin{pmatrix}r_1&r_2&\ldots&r_{\nu}\\j_1&j_2&\ldots&j_{\nu}\end{pmatrix}.
\end{equation}
Now, we split the summation on the right hand side of the above equation into two parts, the first one is $r_1=0$, and the second one is $r_1\geq1$. It is noted that the second one is equal to $\tilde{M}\begin{pmatrix}j_1&j_2&\ldots&j_{\nu}\end{pmatrix}$. Thus, the numerator of the expression in (\ref{d12}) takes the
value
\begin{equation}\label{d25}
\begin{aligned}
\tilde{\Omega}&=\det(I+\tilde{M}+{g^T}E)-\det(I+\tilde{M})\\
&=\sum_{\nu=1}^N\sum_{1\leq{j_1}<{j_2}<\ldots<{j_{\nu}}\leq{N}}\sum_{1\leq{r_2}<\ldots<{r_{\nu}}\leq{N}}
\tilde{K}\begin{pmatrix}j_1&j_2&\ldots&j_{\nu}\\0&r_2&\ldots&r_{\nu}\end{pmatrix}
K\begin{pmatrix}0&r_2&\ldots&r_{\nu}\\j_1&j_2&\ldots&j_{\nu}\end{pmatrix}\\
&=\sum_{\nu=1}^N\sum_{1\leq{j_1}<{j_2}<\ldots<{j_{\nu}}\leq{N}}\sum_{1\leq{r_2}<\ldots<{r_{\nu}}\leq{N}}(-2)^{\nu-1}
{\bar{k}_{r_2}}\ldots{\bar{k}_{r_{\nu}}}{g_{j_1}}{g_{j_2}}\ldots{g_{j_{\nu}}}\\
&\quad\times{h_{r_2}}\ldots{h_{r_{\nu}}}\prod_{l<l^{\prime},m<m^{\prime}}\frac{({k_l}^2-{k_{l^{\prime}}}^2)(k_l-k_{l^{\prime}})
({\bar{k}_m}^2-{\bar{k}_{m^{\prime}}}^2)(\bar{k}_m-\bar{k}_{m^{\prime}})}{({k_l}^2-{\bar{k}_m}^2)(k_l-\bar{k}_m)}.
\end{aligned}
\end{equation}
Hence,
\begin{equation}\label{d26}
u=-\mathrm{i}\frac{\tilde{\Omega}}{\tilde{\mathcal{D}}}.
\end{equation}
By virtue of (\ref{d22}) and Cauchy-Binet formula, similarly, we can calculate the other expressions in (\ref{d12}).
\begin{equation}\label{d27}
\begin{aligned}
\tilde{\mathcal{D}}_1&=\det(I+M)=\det(I+\tilde{M}),\\
\tilde{\Omega}_1&=\det(I+M+{h^T}(\mathcal{G}-\tilde{\mathcal{G}}))-\det(I+M)\\
&=\sum_{\nu=1}^N\sum_{1\leq{j_1}<{j_2}<\ldots<{j_{\nu}}\leq{N}}\sum_{n=1}^N\sum_{1\leq{r_2}<\ldots<{r_{\nu}}\leq{N}}(-1)^{\nu-1}2^{\nu}
{h_{j_1}}{h_{j_2}}\ldots{h_{j_{\nu}}}{g_n}{g_{r_2}}\ldots{g_{r_{\nu}}}\\
&\quad\times{\bar{k}_{j_1}}{\bar{k}_{j_2}}\ldots{\bar{k}_{j_{\nu}}}\prod_{l<l^{\prime},m<m^{\prime}}\frac{({k_l}^2-{k_{l^{\prime}}}^2)(k_l-k_{l^{\prime}})
({\bar{k}_m}^2-{\bar{k}_{m^{\prime}}}^2)(\bar{k}_m-\bar{k}_{m^{\prime}})}{({k_l}^2-{\bar{k}_m}^2)(k_l-\bar{k}_m)},
\end{aligned}
\end{equation}
and
\begin{equation}\label{d29}
\begin{aligned}
&\det(I-N)=\bar{\tilde{\mathcal{D}}}_1,\\
&\det(I-N+{h^T}(\mathcal{G}+\tilde{\mathcal{G}}))-\det(I-N)=-\bar{\tilde\Omega}_1.
\end{aligned}
\end{equation}
From (\ref{d12}), we have
\begin{equation}\label{d30}
v=2{\rm Re}\left(\frac{\tilde{\Omega}_1}{\tilde{\mathcal{D}}_1}\right).
\end{equation}

In particular, for the case of $N=2$, we have
$$\begin{aligned}
\mathcal{D}&=\det(I+\tilde{M})\\
&=1+\frac{2\sigma\bar{k_1}|g_1|^2}{({\bar{k_1}}^2-{k_1}^2)(\bar{k_1}-k_1)}+\frac{2\sigma\bar{k_1}\bar{g_1}g_2}{({\bar{k_1}}^2-{k_2}^2)(\bar{k_1}-k_2)}\\
&\quad+\frac{2\sigma\bar{k_2}g_1\bar{g_2}}{({\bar{k_2}}^2-{k_1}^2)(\bar{k_2}-k_1)}+\frac{2\sigma\bar{k_2}|g_2|^2}{({\bar{k_2}}^2-{k_2}^2)(\bar{k_2}-k_2)}\\
&\quad+4|g_1|^2|g_2|^2\bar{k_1}\bar{k_2}\frac{({\bar{k_1}}^2-{\bar{k_2}}^2)(\bar{k_1}-\bar{k_2})({k_1}^2-{k_2}^2)({k_1}-k_2)}{\mathcal{T}},\\
\Omega&=\det(I+\tilde{M}+{g^T}E)-\det(I+\tilde{M})\\
&=g_1+g_2+2\sigma g_1g_2({k_2}^2-{k_1}^2)(k_2-k_1)\\
&\quad{\times}[\frac{\bar{k_1}\bar{g_1}}{({\bar{k_1}}^2-{k_1}^2)(\bar{k_1}-k_1)({\bar{k_1}}^2-{k_2}^2)(\bar{k_1}-k_2)}\\
&\quad+\frac{\bar{k_2}\bar{g_2}}{({\bar{k_2}}^2-{k_1}^2)(\bar{k_2}-k_1)({\bar{k_2}}^2-{k_2}^2)(\bar{k_2}-k_2)}],
\end{aligned}$$
where
\begin{displaymath}
\mathcal{T}=\prod\limits_{l,m=1}^2({\bar{k_l}}^2-{k_m}^2)(\bar{k_l}-k_m).
\end{displaymath}
Hence, we can obtain
\begin{equation}\label{d15}
u=-\mathrm{i}\frac{\Omega}{\mathcal{D}},
\end{equation}
By calculating it is easy to obtain
$$\begin{aligned}
{\Omega}_1&=\det(I+M+{h^T}(\mathcal{G}-\tilde{\mathcal{G}}))-\det(I+M)\\
&=\frac{2\sigma\bar{k_1}|g_1|^2}{{k_1}^2-{\bar{k_1}^2}}+\frac{2\sigma\bar{k_1}g_2\bar{g_1}}{{k_2}^2-{\bar{k_1}^2}}+\frac{2\sigma\bar{k_2}g_1\bar{g_2}}{{k_1}^2-{\bar{k_2}^2}}
+\frac{2\sigma\bar{k_2}|g_2|^2}{{k_2}^2-{\bar{k_2}^2}}\\
&\quad+4\bar{k_1}\bar{k_2}|g_1|^2|{g_2}|^2\frac{({\bar{k_1}}^2-{\bar{k_2}}^2)(\bar{k_1}-\bar{k_2})({k_1}^2-{k_2}^2)({k_1}-k_2)(\bar{k_1}+\bar{k_2}-k_1-k_2)}
{\mathcal{T}},\\
&\quad\det(I+M)=\mathcal{D},\quad \det(I-N)=\bar{\mathcal{D}},\\
&\quad \det(I-N+{h^T}(\mathcal{G}+\tilde{\mathcal{G}}))-\det(I-N)=-\bar{\Omega}_1.
\end{aligned}$$
Hence, by means of (\ref{d12}), we get
\begin{equation}\label{d17}
v=2{\rm Re}\left(\frac{{\Omega}_1}{{\mathcal{D}}}\right).
\end{equation}

As an illustration, we give the graphic of two-soliton solution for $\sigma=-1$ and $\sigma=1$, respectively.
\begin{figure}[h]
\centering
\includegraphics[width=4.0cm,angle=-90]{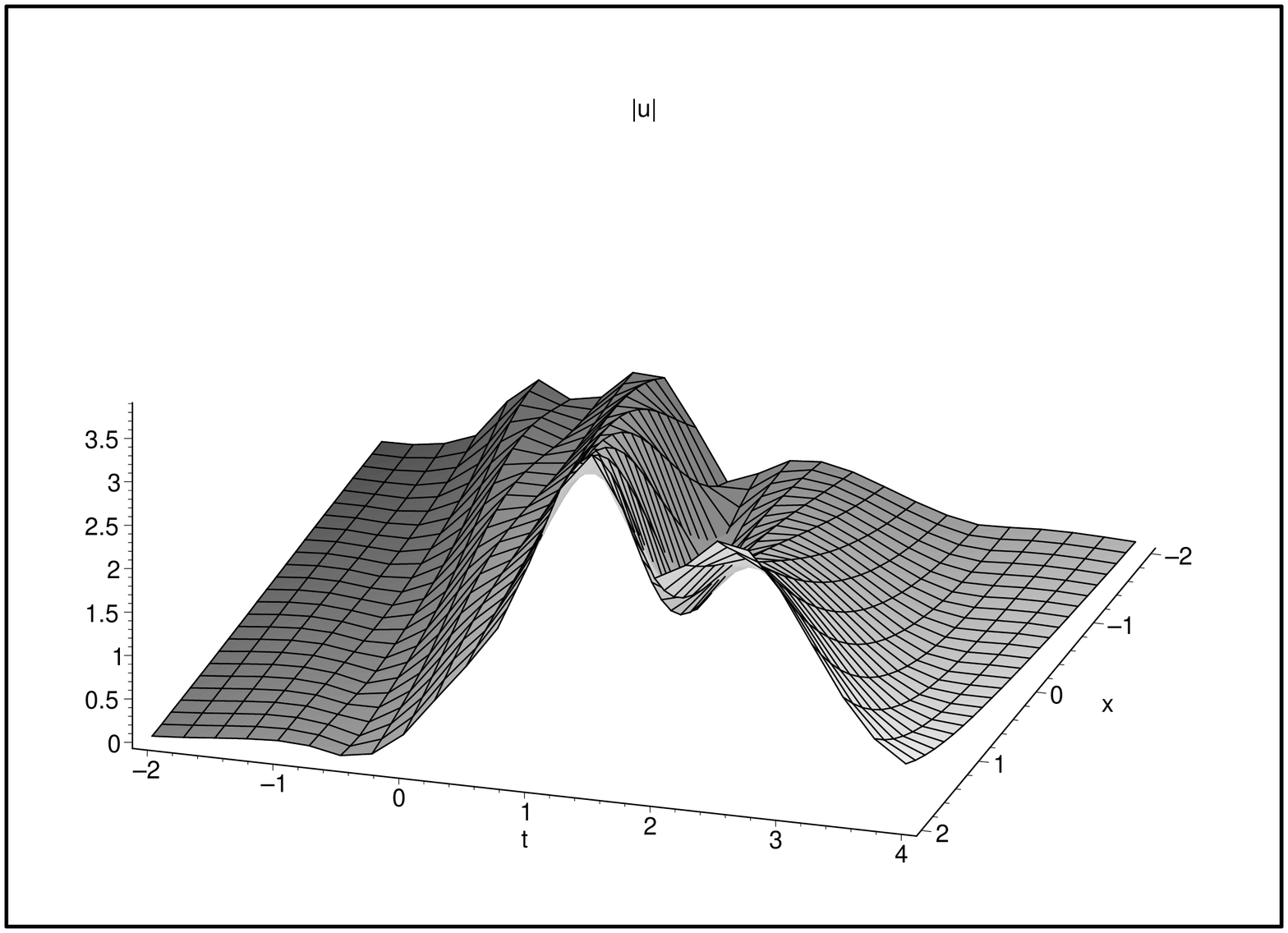}
\includegraphics[width=4.0cm,angle=-90]{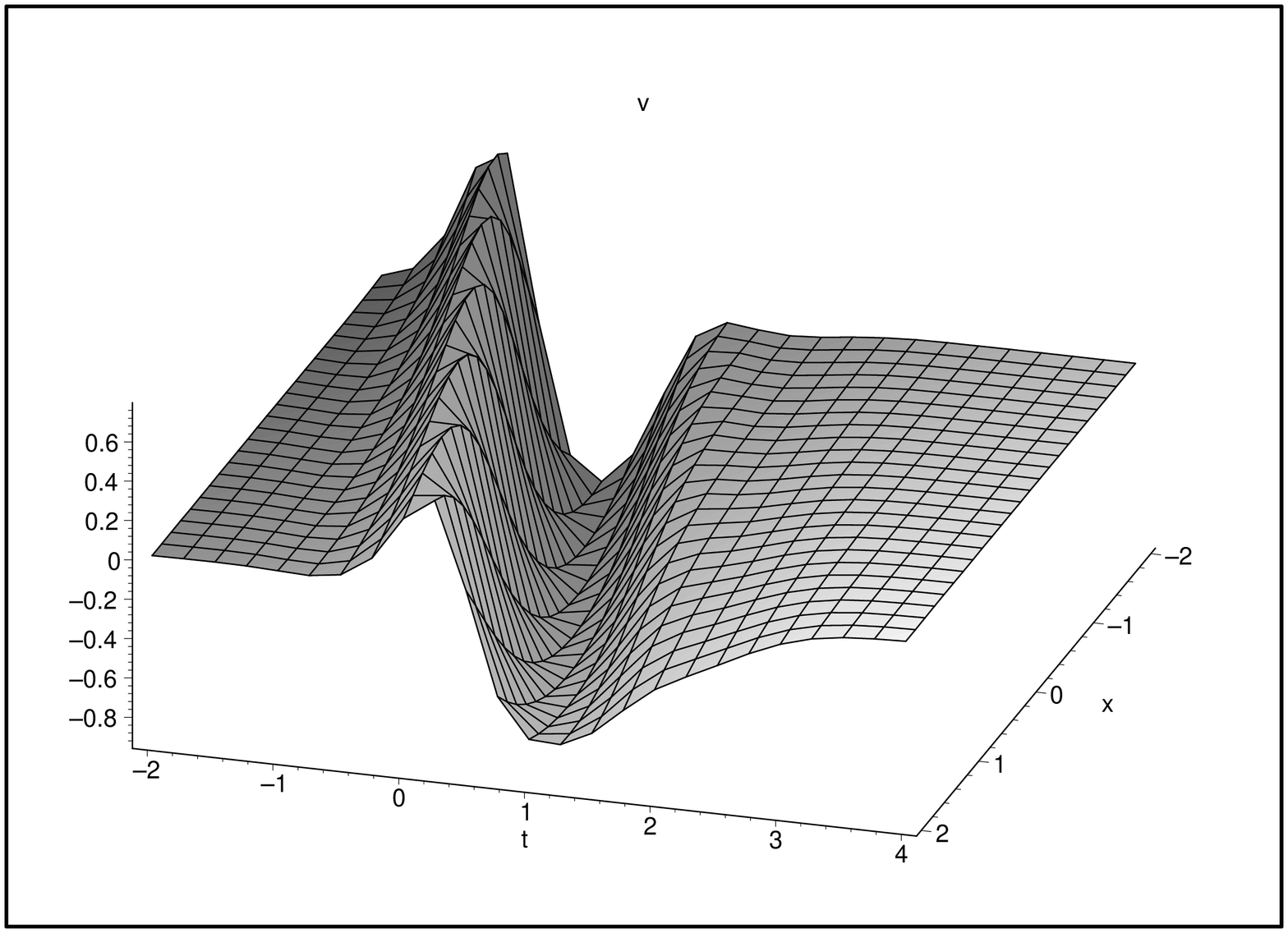}
\caption{$k_1=1.04+0.6\mathrm{i},k_2=2+0.4\mathrm{i},\sigma=-1,z_j=0,\varphi_j=0,j=1,2$.}
\end{figure}
\begin{figure}[h]
\centering
\includegraphics[width=4.0cm,angle=-90]{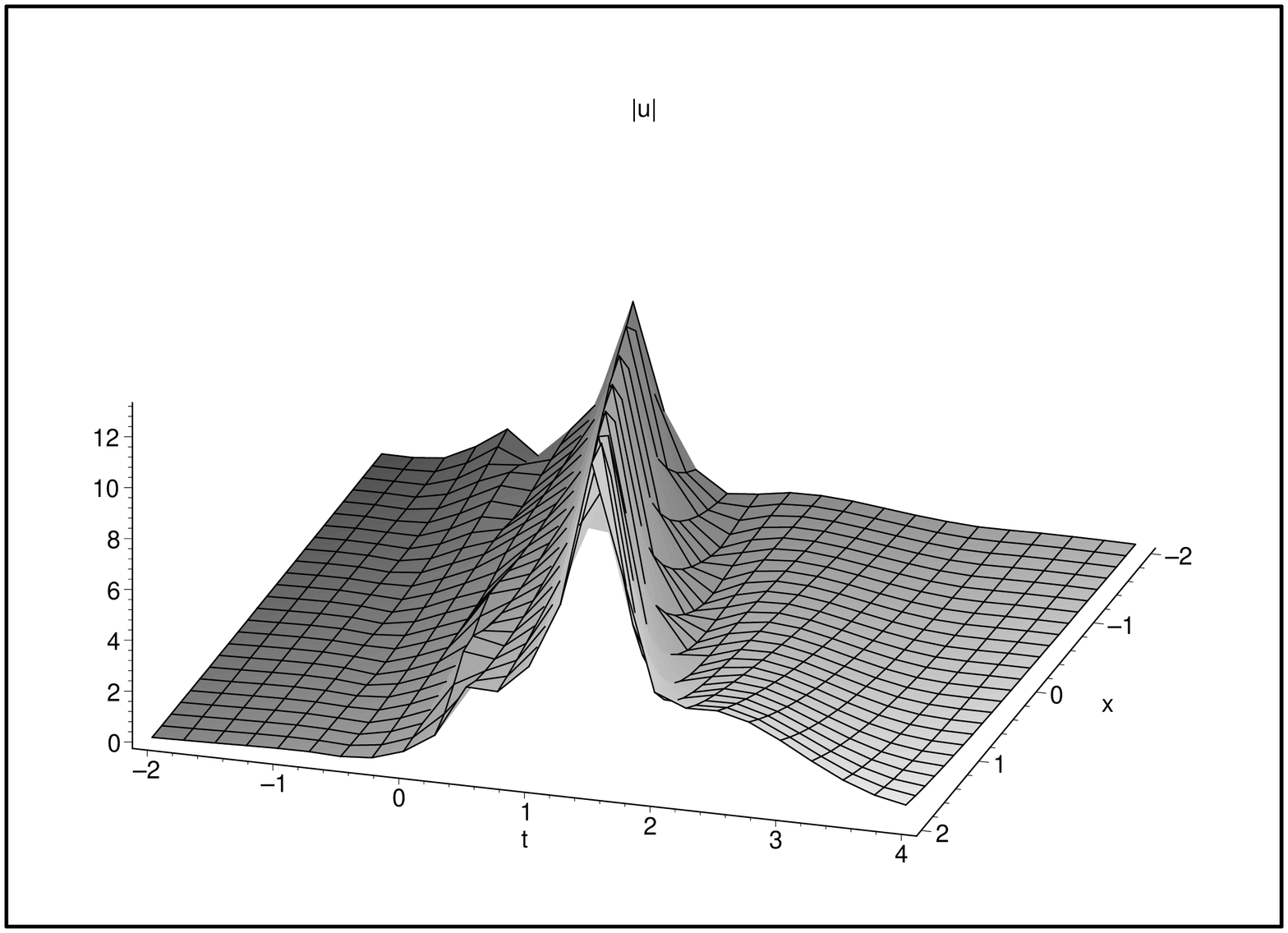}
\includegraphics[width=4.0cm,angle=-90]{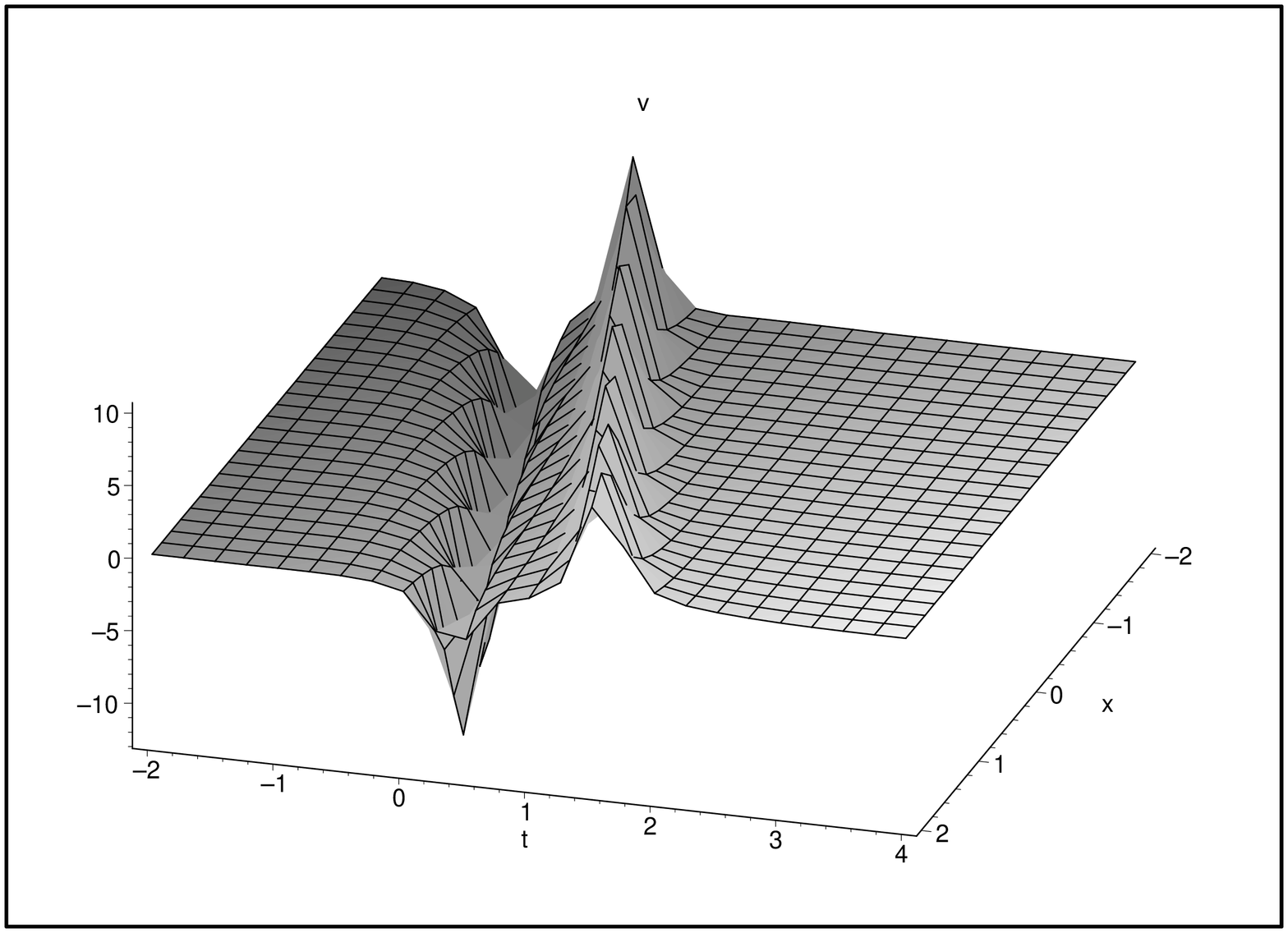}
\caption{$k_1=1.04+0.6\mathrm{i},k_2=2+0.4\mathrm{i},\sigma=1,z_j=0,\varphi_j=0,j=1,2$.}
\end{figure}

It is remarked that from Figures 2 and 4, one-soliton and two-soliton have singularities at the peak of the solitons, which we are called the cusp solitons \cite{W-I-S}.

\setcounter{equation}{0}
\section*{Acknowledgments}
Projects 11301487 and 11171312 are supported by the National Natural Science Foundation of China.
The work of JY Zhu is partially supported by the Foundation for Young Teachers in Colleges and Universities of Henan Province (2013GGJS-010).

\end{document}